\documentclass{article}
\usepackage[english]{babel}
\usepackage[margin=0.5in]{geometry}
\usepackage{multicol}
\usepackage{background}

\usepackage{amsmath,amssymb}	
\usepackage{graphicx}			
\usepackage{here}                             

\SetBgScale{4}
\SetBgColor{gray}
\SetBgAngle{90}
\SetBgContents{Draft}
\SetBgPosition{-1.5,-10}
\begin{document}
 
{\centering
 
{\bfseries\Large Reversal of Helicoidal Twist Handedness Near Point Defects of Confined Chiral Liquid Crystals\bigskip}
 
P. J. Ackerman\textsuperscript{1,2} and I. I. Smalyukh\textsuperscript{1,2,3,4} \\
   {\itshape
\textsuperscript{1}Department of Physics, University of Colorado, Boulder, Colorado 80309, USA \\
\textsuperscript{2}Department of Electrical, Computer and Energy Engineering, University of Colorado, Boulder, Colorado 80309, USA \\
\textsuperscript{3} Soft Materials Research Center and Materials Science and Engineering Program, University of Colorado, Boulder, Colorado 80309, USA \\
\textsuperscript{4} Renewable and Sustainable Energy Institute, National Renewable Energy Laboratory and University of Colorado, Boulder, Colorado 80309, USA \\
\normalfont (Dated: May 9, 2016)
 
   }
}
 
\begin{abstract}
Handedness of the director twist in cholesteric liquid crystals is commonly assumed to be the same throughout the medium, determined solely by the chirality of constituent molecules or chiral additives, albeit distortions of the ground-state helicoidal configuration often arise due to the effects of confinement and external fields. We directly probe the twist directionality of liquid crystal director structures through experimental three-dimensional imaging and numerical minimization of the elastic free energy and show that spatially localized regions of handedness opposite to that of the chiral liquid crystal ground state can arise in the proximity of twisted-soliton-bound topological point defects. In chiral nematic liquid crystal confined to a film that has a thickness less than the cholesteric pitch and perpendicular surface boundary conditions, twisted solitonic structures embedded in a uniform unwound far-field background with chirality-matched handedness locally relieve confinement-imposed frustration and tend to be accompanied by point defects and smaller geometry-required energetically costly regions of opposite twist handedness. We also describe a new spatially localized structure, dubbed a ``twistion'', in which a twisted solitonic three-dimensional director configuration is accompanied by four point defects. We discuss how our findings may impinge on the stability of localized particle-like director field configurations in chiral and non-chiral liquid crystals.
\bigskip
 
\noindent DOI: 10.1103/PhysRevE.93.052702
 
\end{abstract}

\begin{multicols}{2}

\section{Introduction}

A large number of technologically useful and fundamentally interesting effects can arise in liquid crystals (LCs) due to the competition of bulk elasticity and surface anchoring. These effects are especially significant in chiral LCs in which this competition is enriched by the director's tendency to form twisted structures \cite{Chaikin2000}.  For example, perpendicular surface boundary conditions at interfaces of thin chiral nematic (cholesteric) LC films cannot be matched to the cholesteric ground-state helicoidal structure without distortion. This incompatibility results in frustration and both uniform unwound and localized twisted director field configurations \cite{Oswald2000, Smalyukh2010, Chen2013, Ackerman2012, Fukuda2011, Ackerman2014, Ackerman2015NatCom, Varanytsia2015}. This delicate interplay of chirality and confinement enables the spontaneous or laser- and field-assisted realization of many configurations that locally incorporate twist. These geometric-frustration-enabled configurations include at least four different types of localized linear structures dubbed ``cholesteric fingers'' \cite{Oswald2000} and an even larger variety of axially symmetric configurations which include different types of torons, baby-skyrmions, hopfions, and their self-assemblies accompanied by additional defects \cite{Smalyukh2010, Chen2013, Ackerman2012, Fukuda2011, Ackerman2014, Ackerman2015NatCom, Varanytsia2015}. Similar to condensed matter systems such as chiral ferromagnets \cite{Wilson2014, Bogdanov1994jmmm}, the chiral nematic LC's intrinsic tendency to twist helps to stabilize the twisted solitonic structures \cite{Smalyukh2010}. However, we find that the director structures of localized configurations in such confinement-frustrated thin cholesteric films can exhibit regions with winding opposite to that of the twist handedness in the LC's unconfined ground-state \cite{Smalyukh2010}. Furthermore, three-dimensional (3D) imaging of the director field often reveals the 3D spatial extent and geometric shape of small, localized regions within such structures where the director appears to twist in an opposite direction from what is expected based on the handedness of the LC's unconfined ground-state helicoidal structure \cite{Smalyukh2010}. The limited understanding of the physical underpinnings behind such unexpected observations can be attributed to the difficulties in quantitative analysis of the spatial variations of twist handedness and how this twist relates to the free energy cost and topology of the localized field configurations. Since the chiral nature of cholesteric LCs helps to stabilize twisted and knotted solitonic configurations with important analogs from other branches of physics in phenomena that range from elementary particles to cosmology \cite{Ackerman2015NatCom}, knowledge about the handedness reversal due to geometrical confinement is important for enabling LCs as model systems in the exploration of particle-like topological field configurations \cite{Ackerman2015NatCom, Mosseri2008}.

In this work, we show that isotropic and direction-specific handedness of twist in confined cholesteric LCs can reverse within spatially localized regions. We adopt the recently introduced approach of Efrati and Irvine \cite{Efrati2014} to visualize 3D spatial texture of director twist handedness around localized twisted structures that are accompanied by point defects. We then combine this analysis with topological maps of the three-dimensional director field and with spatial distributions of free energy density. When describing our findings below, we use the term ``chirality'' to describe the lack of mirror symmetry per Kelvin's definition \cite{Efrati2014, Kelvin1904} and use the term handedness to describe the property that distinguishes between right- and left-handed director configurations, consistent with the definition of  Efrati and Irvine \cite{Efrati2014}. Using a handedness pseudotensor \cite{Efrati2014}, we characterize the handedness of director twisting as a function of spatial coordinates, the directions along which this twisting occurs, and also the spatial changes of direction-independent ``isotropic handedness''. We show that the experimentally reconstructed director configurations \cite{Trivedi2010} closely match their numerically simulated counterparts which are obtained by minimizing the Frank-Oseen elastic free energy \cite{Chaikin2000} under given perpendicular boundary conditions at confining surfaces. We argue that the small regions of chirality-mismatched handedness arise to enable the confinement of the twisted solitonic structures and their associated point defects surrounded by uniform director into a thin cell. We discuss how our findings may influence the research efforts of stabilization of different localized twisted structures in both non-chiral and chiral LCs. Although a large variety of localized twisted solitonic structures were reported in confined chiral LC systems in previous studies, these field configurations were translationally invariant linear cholesteric fingers or axially symmetric torons, baby skyrmions, and hopfions \cite{Oswald2000, Smalyukh2010, Chen2013, Ackerman2014,Gil1998prl, Smalyukh2005, Porenta2014, Zhang2015}. This work uncovers a new type of localized twisted configuration with a stretched double-twist tori, dubbed ``twistion'', that is accompanied by four self-compensating hyperbolic point defects and small twisted regions with handedness opposite to that of the ground-state cholesteric LC as well as a much larger region of energetically favorable twist within the twistion.

\section{Materials and techniques}

\subsection{Sample preparation}

LC cells were constructed from glass plates of 170 $\mu$m and 1 mm in thickness. The thinner plates of the cells enabled optical imaging with high-numerical-aperture oil-immersion objectives that have short working distances. We spin-coated these substrates with polyimide JALS-204 (JSR, Japan) then cross-linked by baking at about 220 $^\circ$C for one hour to form thin alignment layers that defined strong perpendicular boundary conditions for the director field $\mathbf{n(r)}$ at the inner surfaces of the confinement cell. Alternatively, perpendicular boundary conditions were established by a molecular monolayer created by dip-coating the glass substrates in aqueous solutions of 1 wt.\% [3-(trimethoxysilyl) propyl]octadecyl-dimethylammonium chloride (DMOAP). The LC cell gap thickness was defined using glass spacers or, alternatively, using strips of Mylar film placed along the cell edges and was varied from 5 to 20 $\mu$m. The chiral nematic LC mixtures, that were prepared as described below, were introduced into these cells by capillary forces at an elevated temperature at which the LC transitioned to the isotropic phase. The elevated temperature state prevented flow-induced alignment effects that are associated with filling in the nematic phase. LC mixtures were prepared by doping the nematic host 5CB, ZLI-2806, ZLI-3412, or MCL-6609 (Table I) with a right-handed (CB15) or left-handed (ZLI-811) chiral additive. Control of the volume fraction of the chiral agent allowed us to adjust the equilibrium value of the cholesteric pitch $p$ as described elsewhere \cite{Smalyukh2010}. The partially polymerizable chiral nematic LC was prepared by first mixing 69\% of non-reactive nematic mixture 5CB with 30\% of a diacrylate nematic (consisting of 12\% of RM 82 and 18\% of RM 257 in the final mixture) and 1\% photoinitiator Irgacure 184 (CIBA Specialty Chemicals) which was then followed by doping of this nematic mixture 

\begin{center} 
Table I. Material parameters of the used nematic LC hosts.
\begin{tabular}{lcccc}
\hline \hline
LC$\backslash$Properties & $K_{11},pN$&$K_{22},pN$&$K_{33},pN$&$K_{24},pN$ \\ \hline
5CB              & ~6.4       & 3.0~        & 10.0        & 3.0~      \\
MLC-6609    & 17.2       & 7.51        & 17.9        & 7.51      \\
ZLI-2806      & 14.9       & 7.9~        & 15.4        & 7.9~      \\
ZLI-3412      & 14.1       & 6.7~        & 15.5        & 6.7~      \\
\hline \hline
\end{tabular}
\end{center}

\noindent with chiral additive to obtain the desired cholesteric pitch $p$ \cite{Evans2013}. All LC materials were obtained from EM Industries and used as supplied. The ensuing mixture was first dissolved in dichloromethane to homogenize, heated to 85 $^{\circ}$C for one day to remove the solvent through slow evaporation, and then cooled down to obtain a room-temperature chiral nematic mixture with the equilibrium cholesteric pitch of interest. The use of different nematic hosts and chiral additives allowed us to test how the observed structures of solitons and handedness of twist depend on material parameters such as elastic constants (Table I), allowing us to conclude that the twist handedness reversal effects reported in this work are observable for a broad range of material parameters of chiral nematic LCs (and for all studied here nematic hosts and chiral additives).

\subsection{Optical imaging and director reconstruction}

3D imaging of $\mathbf{n(r)}$-structures was performed with an integrated optical setup built around an inverted optical microscope IX-81 (Olympus) that was capable of simultaneous optical manipulation with laser tweezers, conventional polarizing optical microscopy (POM) and three-photon excitation fluorescence polarizing microscopy (3PEF-PM) studies \cite{Trivedi2010}. The 3PEF-PM approach used a self-fluorescence signal from the LC molecules (detected within the spectral range of 400-450nm), which were excited through three-photon absorption \cite{Trivedi2010}. Laser manipulation used a reflective, electrically addressed, phase-only spatial light modulator (P512-1064, Boulder Nonlinear Systems) with $512 \times 512$ pixels, each $15 \times 15$  $\mu$m$^2$ in size, and an Ytterbium-doped fiber laser (YLR-10-1064, IPG Photonics) that operated at 1064 nm. The trapping beam polarization was controlled with a Glan-laser polarizer and a half-wave retardation plate. The spatial light modulator controlled the phase of the beam on a pixel-by-pixel basis according to computer-generated holograms that were supplied at a refresh rate of 30 Hz. This spatially phase-modulated beam was then imaged at the back aperture of the microscope objective. Such a configuration allowed us to generate the 3D light intensity patterns used for the laser trapping and manipulation \cite{Trivedi2010}. 3PEF-PM imaging employed a tunable (680-1080 nm) Ti-Sapphire oscillator (Chameleon Ultra II, Coherent) as an excitation light source that emitted 140 fs pulses at the repetition rate of 80 MHz. The 3PEF-PM signal from the LC was collected in epi-detection mode after a set of interference filters with a photomultiplier tube (H5784-20, Hamamatsu). We used an oil-immersion 100$\times$ objective with numerical aperture NA=1.4 (Olympus). The vertical position of the focal point within the sample was adjusted with a stepper motor (10 nm precision). 3D images were reconstructed based on processing 3PEF-PM self-fluorescence signals that originated from individual pixels with ImageJ (National Institute of Health), FluoView (Olympus), and ParaView (Kitware Inc.) software. The effects of defocusing and polarization changes due to the medium's birefringence were mitigated by the use of LCs with low birefringence or the partial photopolymerization approach that allowed for an order-of magnitude reduction of the effective birefringence upon the replacement of the unpolymerized component of the system with an immersion oil \cite{Evans2013}. The integrated setup, the details of its operation, and the different imaging approaches are described in detail elsewhere \cite{Trivedi2010, Evans2013}. Director structures were reconstructed from multiple 3D 3PEF-PM images that were obtained with different polarizations of the excitation light. The coordinate-dependent angles $\alpha\mathbf{(r)}$ between $\mathbf{n(r)}$ and the linear polarization of the probing laser light were determined based on the 3PEF-PM signal dependence $I_{3PEF-PM} \propto cos^n\alpha$ (where $n=6$ for the case of unpolarized self-fluorescence detection that was used in this work) and a series of images for different polarizations of excitation light \cite{Trivedi2010}. To distinguish between two possibilities of the director tilt within the structures, $\alpha$ versus $-\alpha$, additional cross-sectional test images were also obtained for the tilted orientations of the LC cell normal at $\pm2^{\circ}$ with respect to the microscope's optical axis when linear polarization of excitation of laser light was set to be parallel or perpendicular to the plane of the corresponding 3PEF-PM vertical cross-section. These tests allowed us to eliminate the possible tilt ambiguity in regard to reconstructed director structures based on the $\propto cos^6\alpha$ angular dependence and the decrease or increase of the 3PEF-PM signal upon the $\pm2^{\circ}$ sample tilting.

Experimental reconstruction of 3D $\mathbf{n(r)}$-structures was based on analyzing the spatial variations in the 3D images of 3PEF-PM intensity $I_{3PEF-PM}$ data \cite{Chen2013, Trivedi2010, Evans2013}. For the reconstruction of solitonic structures, we defined the director in terms of polar $\theta$ and azimuthal $\phi$ angles, $\mathbf{n(r)}=[sin\theta cos\phi,~sin\theta sin\phi,~cos\theta]$. The angle $\alpha$ between the in-plane direction of the linear polarization of excitation light and $\mathbf{n(r)}$ was then expressed in terms of $\phi$. Since we knew from the experiments that $\theta=0$ (the director was normal to the top and bottom surfaces) far from the studied localized structure in the used homeotropic cells, we normalized the 3PEF-PM intensity $I_{3PEF-PM}$ to take values within 0-1. Using four 3D 3PEF-PM images with the probing laser light linearly polarized at four different angles $\pi/4$ apart in the x-y plane $(I_0,~I_{\pi/4},~I_{\pi/2},~I_{3\pi/4})$, we then calculated the Stokes parameters $S_0$, $S_1$, and $S_2$ as follows:

\begin{equation}
S_0 =  \frac{1}{2}(I_0+I_{\pi/4}+I_{~\pi/2}+I_{3\pi/4}) 
\end{equation}
\begin{equation}
S_1 =  I_0-I_{~\pi/2} 
\end{equation}
\begin{equation}
S_2 =  I_{\pi/4}-I_{3\pi/4}
\end{equation}

We then used these parameters to calculate coordinate-dependent $\theta(\bf{r})$ and $\phi(\bf{r})$ by exploiting the corresponding relations $S_0 \propto J \sin^6 (\theta)$ and $S_1/S_2=\tan(2\phi)$, where $J$ is the amplitude of the normalized 3PEF-PM signal. The reconstructed $\mathbf{n(r)}$ is determined from the series of four images and additional images for two small tilts of the sample normal with respect to the microscope's optical axis used to eliminate the ambiguity in determining the director tilt directionality as discussed above. The ensuing 3D $\mathbf{n(r)}$-configurations were visualized in Paraview with the help of a series of nested two-dimensional iso-surfaces of constant $\theta$, or equivalently iso-surfaces of constant z-component of the LC director (here the z-axis is aligned along the cell normal), that are color-coded by $\phi$ \cite{Chen2013}. Each of such surfaces captures topological information about the studied 3D $\mathbf{n(r)}$-field. Additionally, through the use of a series of nested iso-surfaces of different constant $\theta$ and with color-coded $\phi$ we visually present not just topology but also many other important details of the 3D $\mathbf{n(r)}$-configurations that we discuss below. 

\subsection{Modeling through free energy minimization and handedness analysis}

In addition to the reconstruction from experiments, the 3D $\mathbf{n(r)}$-configurations were also obtained by means of numerical minimization of the Frank-Oseen elastic free energy 

\begin{multline}
F_{elastic} = \int\bigg \{\frac{k_{11}}{2}(\nabla \cdot \mathbf n)^2+\frac{k_{22}}{2}(\mathbf n \cdot \nabla \times \mathbf n + q_0)^2\\+\frac{k_{33}}{2}(\mathbf n \times \nabla \times \mathbf n)^2\\  
-k_{24}\nabla \cdot (\mathbf n \nabla \times \mathbf n + \mathbf n \times \nabla \times \mathbf n)\bigg \}dV
\label{eq:fe_handedness}
\end{multline}
Where $k_{11},~k_{22},~k_{33},~k_{24}$ are elastic constants that pertain to splay, twist, bend, and saddle-splay distortions of $\mathbf{n(r)}$, respectively. Values for elastic constants that were used in experiments are summarized in Table I. The precise values of $k_{24}$ were unknown for the studied materials, so it was assumed that $k_{24}=k_{22}$, as in the previous studies \cite{Smalyukh2010}. The minimization was performed with the director relaxation method \cite{Ackerman2015NatCom}. All simulations were performed for a volume of $26 \times 26 \times 8$ $\mu$m that was split into $104 \times 104 \times 32$ equally spaced voxels on a 3D grid. The free-energy-minimizing computer-simulated field configurations were analyzed and compared to experiments with the generation of 3D iso-surfaces that were constructed similar to the experimental ones discussed above. Additionally, we constructed 3D and cross-sectional iso-surfaces and density plots of free energy along with iso-surfaces and cross-sectional plots revealing LC twist handedness, which we describe below.

As introduced by Irvine and Efrati, the components of the handedness pseudotensor $\chi_{ij}$ for the 3D unit vector and director fields, such as $\mathbf{n(r)}$, twisting around a certain unit vector $\mathbf{b}$ when displaced along a direction $\mathbf{a}$ can be expressed as

\begin{equation}
a^i \chi_{ij} b^j = a^i \partial_i n^k \epsilon_{jlk} n^l b^j 
\label{eq:ht}
\end{equation}

where $\epsilon_{jlk}$ is the anti-symmetric Levi-Civita tensor and we use Einstein's summation notation. This measure yields the degree of handedness and associates the handed phenomena with a direction. The trace of the handedness pseudotensor is the isotropic handedness that is reminiscent of physical quantities such as magnetic and hydrodynamic helicities \cite{Efrati2014}:

\begin{equation}
\chi^{ii} = \partial_i n^k \epsilon_{ilk} n^l = -(\nabla \times \mathbf n)\cdot \mathbf n = \mathcal{H}
\label{eq:trht}
\end{equation}

Since $\mathcal{H}$ is quadratic in $\mathbf{n}$, it remains unchanged under the transformation $\mathbf{n}\rightarrow-\mathbf{n}$. This property is consistent with the nonpolar symmetry of the chiral nematic LC. The analyses of 3D spatial patterns of   and the $\chi_{ij}$-components that describe handedness in different directions complement a similar analysis of 3D field configurations and their energetic costs. Both approaches of analyzing different solitons allow us to obtain insights into the physical underpinnings behind the stability of these localized field configurations in confined chiral LCs. 

\section{Results and discussion}

The characterization of handedness and chirality of ground-state cholesteric LCs, as well as of structures that may be considered to be weakly deformed variants of this ground state, was commonly assumed to be a trivial task. The so-called ``right-hand rule'' characterizes the handedness of twist around a single helicoidal axis, determining whether this twist is left-handed or right-handed. Its extension works well also for the double-twist variants of configurations in LCs that are found in cholesteric blue phases and in two-dimensional solitons called ``baby skyrmions''. However, configurations with triple twist and different amounts of twisting around all three Cartesian coordinate axes cannot be easily characterized with only the conventional concepts of chirality and isotropic handedness. We therefore adopt the pseudotensor description of handedness approach of Efrati and Irvine. From the isotropic average of the handedness pseudotensor, we recover a pseudoscalar $\mathcal{H}$ (the trace of the pseudotensor) that is consistent with the Kelvin's definition \cite{Kelvin1904}. 

\begin{figure}[H]
\includegraphics[width=0.49\textwidth]{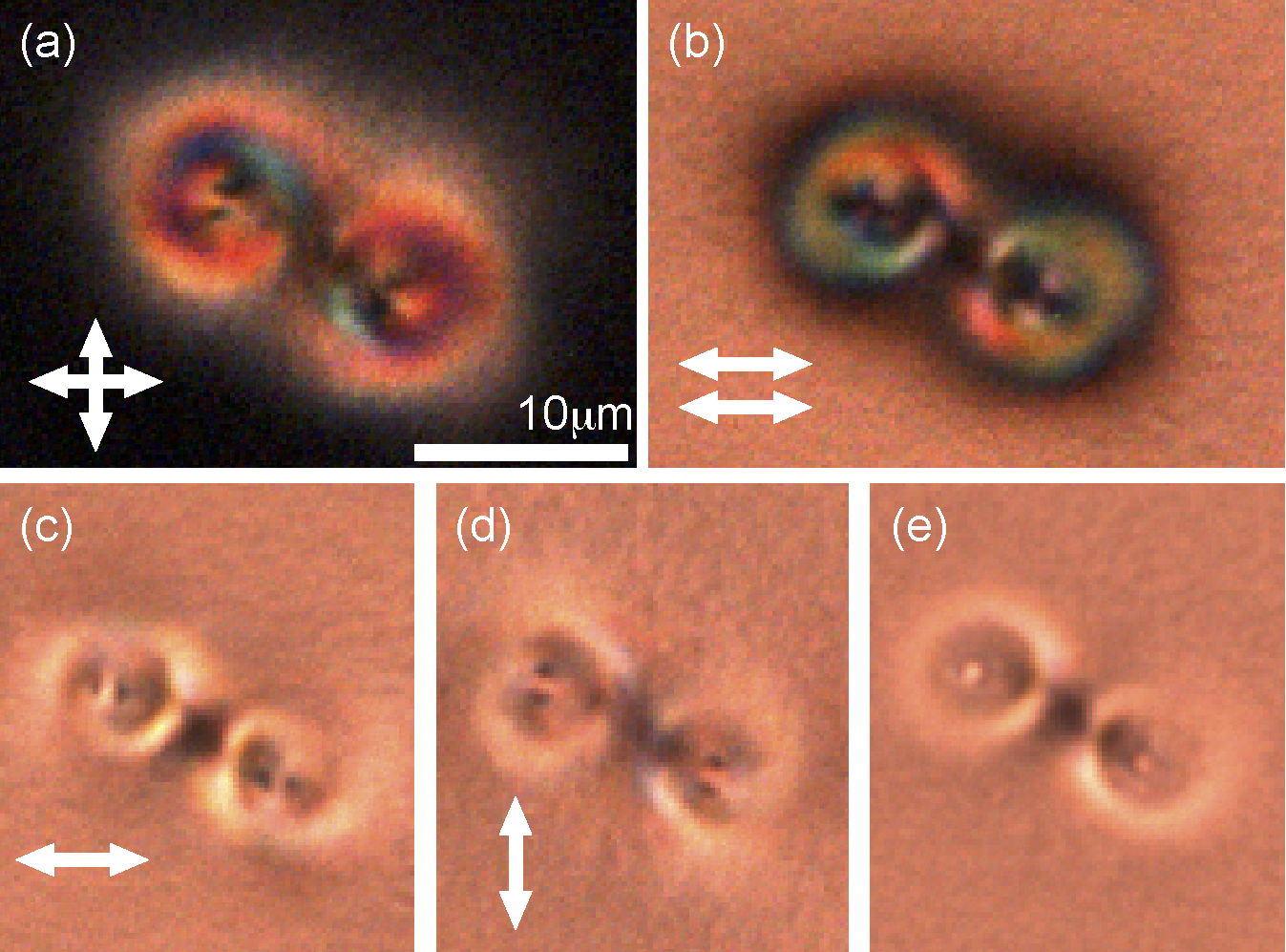}
\caption{Experimental polarizing optical micrographs of a stable
twisted configuration dubbed twistion in a partially polymerizable confined chiral nematic LC as viewed (a) between crossed polarizers, (b) between horizontally oriented parallel polarizers, (c) with only one horizontally and (d) one vertically oriented polarizer, and (e) without polarizers. Double-headed white arrows depict orientations of linear polarization of light passing through these polarizers. The scale bar shown in part (a) applies to all micrographs (a$-$e).}
\label{handednessFig1}
\end{figure}

To illustrate the complex topology and nontrivial spatial variations of director twisting in confined cholesteric LCs, we choose to analyze an example of a localized twisted and stable configuration that we call a ``twistion'' (Fig. \ref{handednessFig1}). When it is

\begin{figure}[H]
\includegraphics[width=0.49\textwidth]{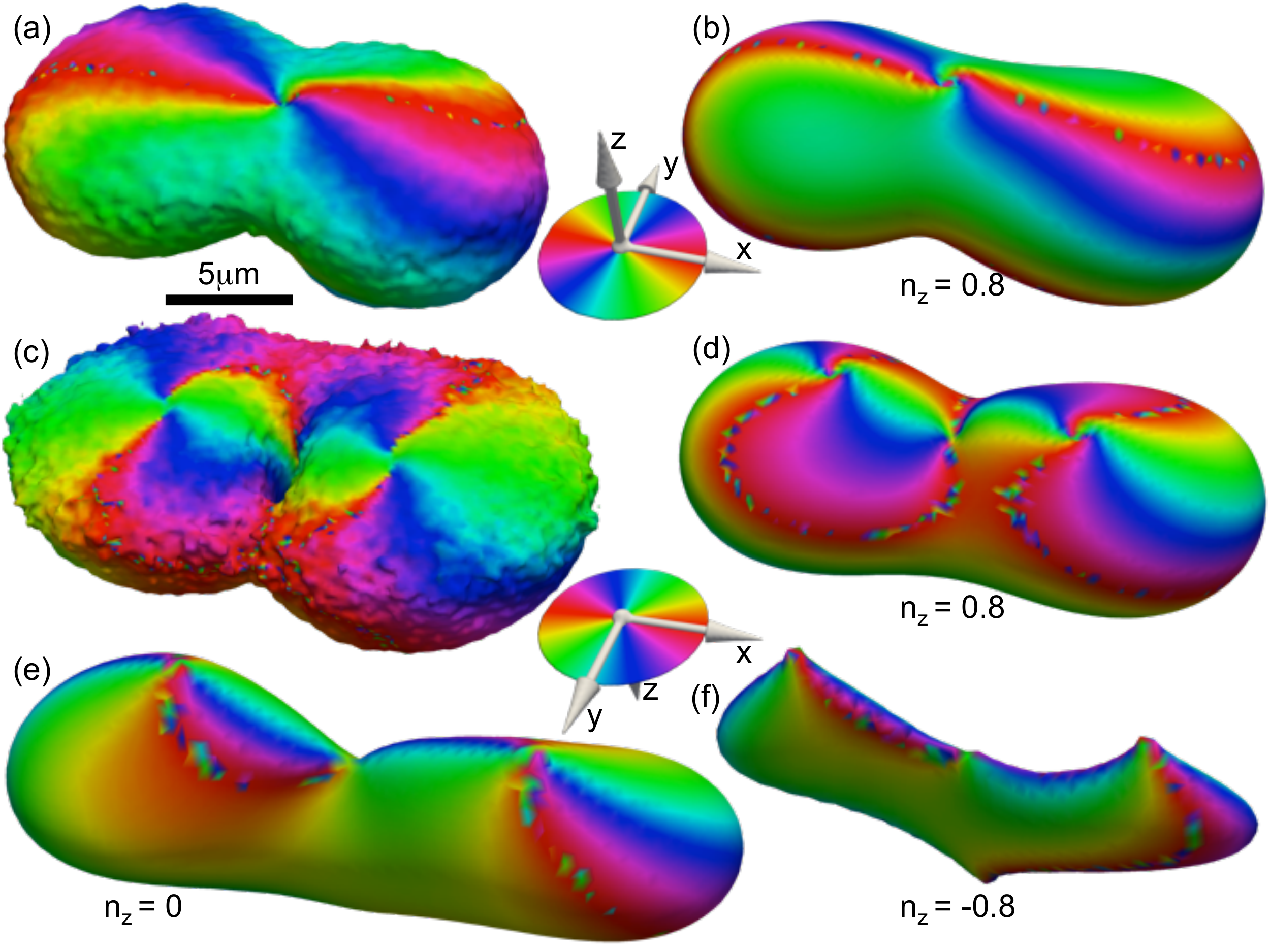}
\caption{3D colored surfaces depicting the twisted director configuration of twistions that were reconstructed from experimental images and numerically simulated data. (a) Experimentally reconstructed and (b) the corresponding computer simulated iso-surface that depicts the director field configuration at constant $n_z = 0.8$, with the azimuthal orientations shown using the color-coded scheme in the inset. The 3D coordinate system in the inset attached to the color scheme helps to describe the orientation of surfaces relative to the viewing perspective direction. The colors visualize azimuthal director orientations that are characterized by the projection of the unit director field to the x-y plane, according to the color wheel that is shown with the coordinate system. (c) An experimental configuration shown for $n_z=0.8$ iso-surface similar to (a) but when viewed from a $-$z perspective view position, as indicated by the flipped coordinate system in the inset next to it. (d-f) Color-coded iso-surfaces visualizing $\bf n(r)$ for (d) $n_z=0.8$, (e) $n_z=0$, and (f) $n_z=-0.8$ constructed from the numerically simulated director field and viewed from the $-$z perspective viewpoint. The scale bar between (a) and (c) relates to experimental 3D reconstructions of the director structures on iso-surfaces shown in (a,c). All experimental images were obtained for the partially polymerized chiral nematic LC system with the composition detailed in the methods section and 5CB elastic constants were used in the corresponding computer simulations.}
\label{handednessFig2}
\end{figure}

\begin{figure}[H]
\begin{center}
\includegraphics[width=0.6\textwidth]{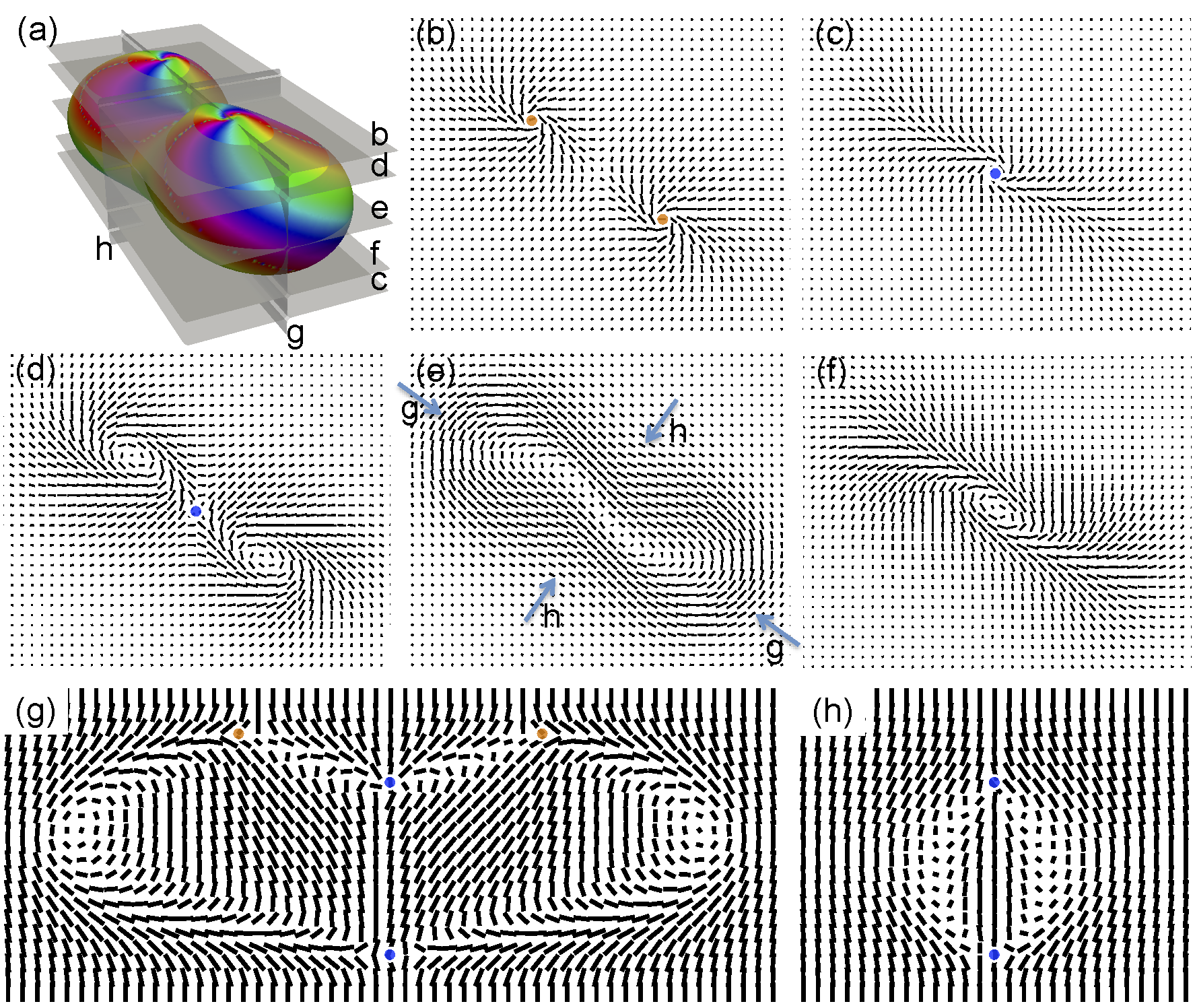}
\end{center}
\caption{Cross-sectional views of the director field within the twistion configurations with point defects. (a) A color-coded iso-surface at $n_z = 0.8$ that was constructed from computer-simulated director field data, shown along with the cross-sectional planes that denote locations of the lateral x-y cross-sections (b) near the top of the structure, (c) near the bottom, (d-f) in-between, and the two mutually orthogonal vertical cross-sections (g,h) that contain point defects. Director field $\mathbf{n(r)}$ local orientations are shown with rods. Locations of the four self-compensating point defects are depicted with orange and blue dots that represent clockwise RGB and BGR color changes of azimuthal orientation of $\mathbf{n(r)}$ around defects for the iso-surface shown in (a), respectively. (g,h) Two mutually orthogonal x-z cross-sections (g) through all four point defects and (h) y-z cross-section through only two point defects shown with the help of enlarged rods depicting $\mathbf{n(r)}$ for clarity. Arrows in (e) mark the locations of these vertical cross-sections, in addition to the planes shown in (a). These results of computer simulations are shown for elastic constants of 5CB.}
\label{handednessFig3}
\end{figure}

\noindent viewed under an optical polarizing microscope, the appearance of this structure resembles that of two torons bound together. The interaction of polarized light with the twistion when it is imaged at different polarizer and analyzer orientations suggests the presence of a localized region with the director twist, which is embedded within the uniform homeotropic far-field and accompanied by four point defects. To understand the topology of this configuration, we use 3D director images to experimentally reconstruct iso-surfaces of constant z-component of $\mathbf{n(r)}$, such as the ones shown in Fig. \ref{handednessFig2}(a,c),

\noindent with the azimuthal orientations of the director represented using colors. Since the point defects of the twistion structure contain all possible orientations of the director around them, these iso-surfaces always embed all four point defects. At the point defects, all colors, which represent different azimuthal director orientations, meet and behave discontinuously. A circumnavigation of the defect on the iso-surface yields sequential color change that is indicative of the relative $\pm$1 hedgehog topological charge of the point defect. When the defect is inspected from the same viewing direction (e.g. along the positive z-axis direction), the change of colors around two defects is found to be in the clockwise direction while that of the other two is in the counterclockwise direction. A comparison of directionality between defects for a given viewing angle indicates that the relative net total charge of the point defects is zero, as is required by charge conservation to embed in the uniform homeotropic background of the cell. Upon reversal of the viewing direction to be along the $-$z-axis, the direction of the color wheel reverses too, consistent with the LC's nonpolar symmetry and the fact that only relative charges $\pm$1 can be assigned to the nematic point defects \cite{Chaikin2000}. The family of iso-surfaces with different constant values of the z-component of $\mathbf{n(r)}$ provides a generalization of the Pontryagin-Thom construction \cite{Chen2013}, that reveals structural details in addition to the topology information. The color pattern changes in accord with the continuous changes of the z-component of $\mathbf{n(r)}$ ($n_z$) and reveals twist of the director field along different spatial directions, albeit, quantification of this director twisting is difficult to achieve with information from such surfaces alone. However, since the color-coded patterns of $\mathbf{n(r)}$ on all examined iso-surfaces closely match the computer-simulated counterparts (Fig. \ref{handednessFig2}), we can use detailed cross-sectional presentations of the computer-simulated 3D director field to examine twisting within these structures (Fig. \ref{handednessFig3}). For this purpose, we choose a series of vertical and horizontal cross-sectional planes that reveal the salient details of the twistion configuration. Consistent with experiments, the simulated structure reveals four hyperbolic point defects. Two of them are located close to one of the confining surfaces while the other two are in the middle of a stretched torus of twist (Fig. \ref{handednessFig3}). The twisted region is comprised of a looped double-twist tube that is stretched in one lateral direction. The depth location of the axis of the double-twist tube changes when circumnavigating it along the closed extended loop it forms, with its locations shifted upwards with respect to the cell midplane in the center of the twistion but downwards at the opposite ends of the twistion (Fig. \ref{handednessFig3}g,h), or vice versa.

\begin{figure}[H]
\begin{center}
\includegraphics[width=0.6\textwidth]{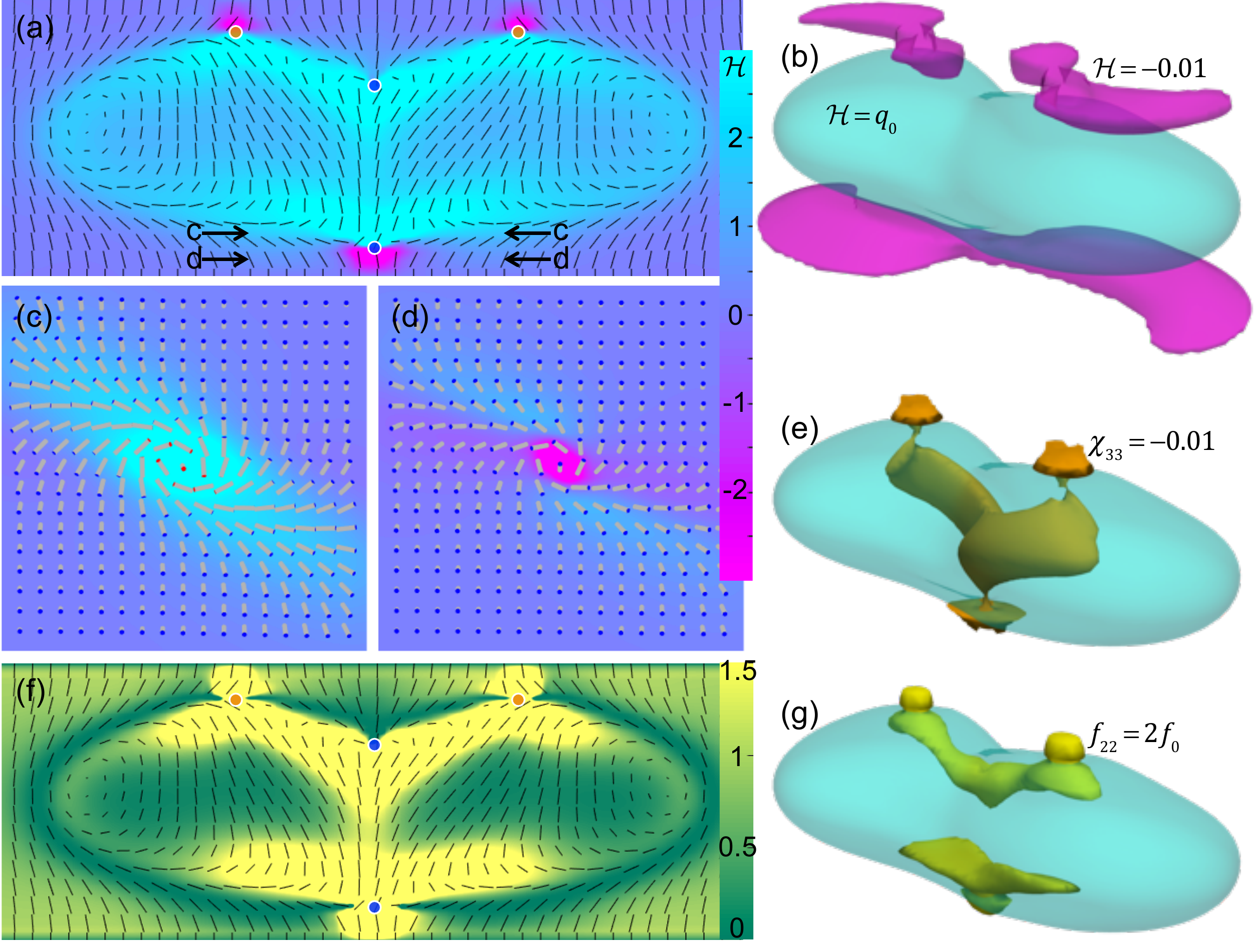}
\end{center}
\caption{Visualization of twist handedness and energy density of twistions in confined chiral LCs. (a) Vertical x-z cross-section of the computer-simulated director field of the twistion (with the director orientations shown using rods) is overlaid on top of a density plot of isotropic handedness normalized to q0 and shown in accord with the color scheme provided in the inset. (b) A 3D perspective view of the iso-surfaces of isotropic handedness that correspond to energetically favorable twist ($\mathcal{H}=q_0$, cyan) and the iso-surface that encloses the volume with opposite handed twist ($\mathcal{H}=-0.01$, magenta). (c,d) In-plane x-y cross-sections of the director structure that are overlaid on top of the density plots of isotropic handedness (c) above and (d) below the bottom point defect, as marked on (a). The director orientation is represented by cylinders with red- and blue-colored ends to help guide the reader's eye when he or she inspects the directional handedness in the corresponding regions. (e) Iso-surface $\chi_{33} = -0.01$ for the twist along z (depicted in orange color) that is shown along with the same cyan iso-surface as in (b) for reference. The reader should note here that the spatial overlap of the two 3D surfaces gives appearance of green color. (f) Vertical x-z director cross-section of the twistion that is overlaid on top of a density plot of the twist term of the free energy density $f_{22}$ normalized to the energy density of the untwisted state f0 and shown with the color scheme that is provided in the right-side inset. (g) Twist free energy iso-surface $f_{22} = 2f_0$ (yellow) shown along with the same handedness iso-surface ($\mathcal{H}=q_0$, cyan) as in (b) for reference. Overlap of the two surfaces again gives the appearance of green. Computer simulations were done for the material parameters for 5CB (Table I).}
\label{handednessFig4}
\end{figure}

To quantitatively probe the details of director twisting within the twistions, we characterize both directional and isotropic handedness as functions of spatial coordinates (Fig. \ref{handednessFig4}). The spatially localized configuration of the twistion has a well-defined region with the sign and magnitude of 

\end{multicols}
\begin{multicols}{2}

\noindent isotropic handedness close to that of the equilibrium-state chiral LC $q_0=2\pi/p$, where $p$ is the equilibrium cholesteric pitch (Fig. \ref{handednessFig4}a,b). The unwound state with uniform homeotropic texture of the confined chiral LC surrounds the twistion, both in the lateral directions and above and below it, where the director meets strong perpendicular boundary conditions. Four hyperbolic point defects help to match the twisted localized structure with the uniform far-field director. Interestingly, the twist rate, which is described by the magnitude of the isotropic handedness, can exceed the equilibrium value and, even more unexpectedly, can also reverse sign within small, localized regions in the vicinity of hyperbolic defects (Fig. \ref{handednessFig4}a-d). The twist handedness reversal takes place in relatively small spatially localized regions close to substrates. In addition to analyzing density plots of isotropic (described by the trace $\mathcal{H}$ of the handedness pseudotensor) and directional handedness (which in our coordinate system is quantified by the diagonal elements $\chi_{11}$, $\chi_{22}$, and $\chi_{33}$ of the handedness pseudotensor), the reversal of handedness can be examined in a conventional way by applying the right-hand rule for certain regions. For example, in the location close to the point defect depicted in Fig. \ref{handednessFig4}d, the magenta-colored region has handedness opposite from that of the equilibrium state when probed along both the horizontal and vertical axes. $\mathcal{H}$ and elements $\chi_{11}$, $\chi_{22}$, and $\chi_{33}$ can have signs opposite from that of the ground-state in different spatial regions, albeit there is also a significant spatial overlap between such regions (Fig. \ref{handednessFig4}a-e). As expected, the spatial regions with opposite twist and different twisting rate significantly differ from that of the ground-state chiral nematic LC and are costly in terms of the free energy. The correlation between the isotropic twisting handedness and local free energy density can be clearly seen from the analysis of density and iso-surface plots of the twist term of the elastic free energy (Fig. \ref{handednessFig4}f,g). The high energy density that corresponds to these regions with over-twisted or oppositely twisted structures supports the notion that the existence of such regions is determined by the need to match optimally twisted regions of the twistion's interior to the unwound LC that is enforced by the vertical surface boundary conditions and the uniform far-field director far away from the localized structures.

\end{multicols}
\begin{multicols}{2}

\begin{figure}[H]
\begin{center}
\includegraphics[width=0.6\textwidth]{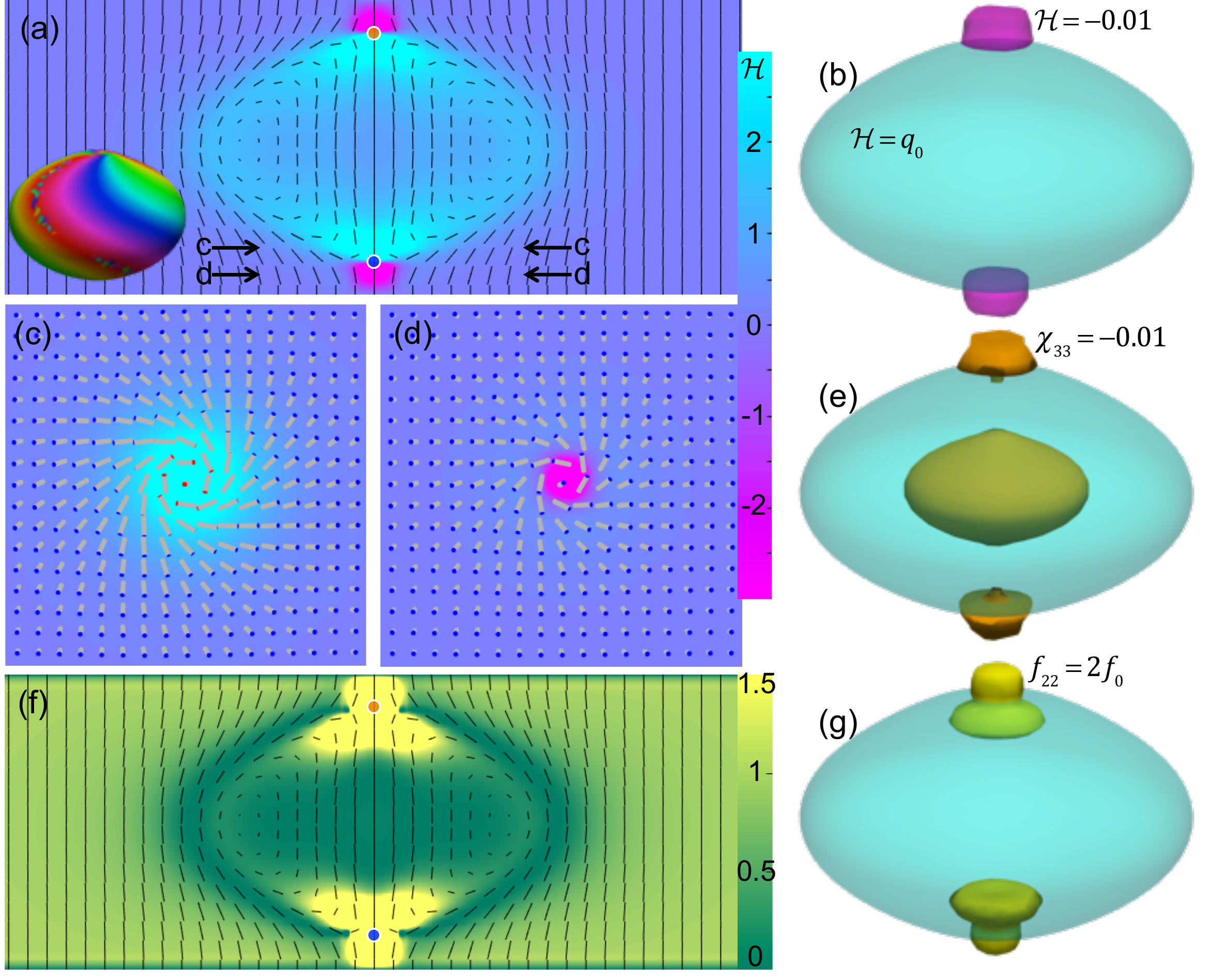}
\end{center}
\caption{Visualization of twist handedness and free energy density within a triple twist toron of the first kind. (a) Vertical x-z director field cross-section of a computer-simulated toron structure that is overlaid on top of a density plot of isotropic handedness normalized to $q_0$, and plotted in accord with the color scheme shown in the right-side inset. The bottom-left inset shows an iso-surface of the 3D director structure for $n_z = 0.5$ of the corresponding computer-simulated toron configuration, where colors depict azimuthal director orientations in accord with a similar color scheme like that used in Fig. \ref{handednessFig2}. (b) 3D perspective view of iso-surfaces in isotropic handedness corresponding to ($\mathcal{H}=q_0$, cyan) and ($\mathcal{H}=-0.01$, magenta). (c,d) In-plane x-y director cross-sections that are overlaid on top of the corresponding density plots of isotropic handedness for cross-sections (c) above and (d) below the bottom point defect, as marked on (a). The local director orientation is represented by cylinders with red- and blue-colored ends to help guide the reader's eye when he or she inspects the directional handedness. (e) Analysis of the 3D patterns of directional handedness represented using orange iso-surface with $\chi_{33} = -0.01$, for the director twist along z, depicted along with the same cyan iso-surface as shown in (b) for reference. The spatial overlap of the two 3D surfaces appears green. (f) x-z director cross-section of the twistion overlaid on top of a density plot of twist free energy density $f_{22}$ normalized by the energy density of the untwisted state $f_0$; the color-coded free energy density scale is shown in the right-side inset. (g) A 3D perspective view on the yellow/green iso-surface with $f_{22} = 2f_0$, shown in accord with the cross-sectional plot depicted in (f). This isosurface is shown along with the same cyan iso-surface of handedness as in (b) for reference. The yellow iso-surface corresponds to the high-energy $f_{22} = 2f_0$ in the regions with the twist direction opposite from that of the ground-state cholesteric LC and for the same handedness but with the twist rate higher than that of the equilibrium helicoidal structure. Overlap of the two surfaces appears green. Computer simulations are based on material parameters of 5CB (Table I).}
\label{handednessFig5}
\end{figure}

Although we used the localized structure of the twistion to demonstrate local reversal of both isotropic and directional handedness, this behavior is common for many other localized twisted structures in confined chiral nematic LCs. For example, the simplest type of the triple twist toron (Fig. \ref{handednessFig5}) with two point defects separated by a torus of double twist also features near-defect regions with the handedness reversal. Similar to the case of twistions, director twisting within this axially symmetric structure of toron is observed both along the cell normal and in the lateral directions, but with oppositely twisted regions in-between the point defects and the confining substrates. The representation of the director field with the help of rods with differently colored ends enables analysis of this handedness twist reversal in the planes below and above the point defects (Fig. \ref{handednessFig5}c,d). Such a scheme makes the twisting reversal within the magenta-colored regions apparent even when one applies the conventional right-hand rule, albeit the iso-surfaces of isotropic and directional handedness reveal the nature of this phenomenon more quantitatively (Fig. \ref{handednessFig5}b,e). Similar to the case of twistions, the large-volume spatial regions of optimal handedness and twisting rate are the regions where elastic energy is minimized while the oppositely twisted regions and over-twisted regions are costly in terms of the twist term of free energy. Although the director twisting with handedness opposite from the ground-state was noticed previously in both experiments and in numerical modeling \cite{Smalyukh2010}, it could not be quantified and its physical origin could not be easily analyzed. The handedness tensor description of Efrati and Irvine provided a simple framework for the analysis and understanding of this behavior. Similar to twistions, the oppositely handed twist in small regions near toron point defects exist to match the director field within an optimally

\end{multicols}
\begin{multicols}{2}

\noindent twisted localized structure to that of a uniform far-field background. Although the example provided here involves a toron with the net total twist of $\pi$ from the central axis to periphery, with the double-twist torus of the twistion that has a Hopf index of zero, similar studies can be extended to torons and twistions with double twist tori of a larger amount of twist and non-zero Hopf charge \cite{Ackerman2015NatCom} that also possess two hyperbolic defects, which will be described elsewhere. Beyond twistions and torons, we have observed similar types of behavior for different hopfions, cholesteric fingers and finger loops, as well as for other localized structures in confined cholesteric LCs. We will discuss these other localized structures elsewhere.  

We have demonstrated previously that torons can correspond to both equilibrium (global free energy minimum) and metastable (local free energy minima) states of confined cholesteric LCs \cite{Smalyukh2010}, depending on material and geometric parameters such as the cell thickness to pitch ratio, elastic constants, and so on \cite{Smalyukh2010}. It is therefore of interest to compare the free energy of the new twistion configuration to that of the unwound and toron states. We have performed such an analysis and summarize it in Table II. The total free energies, as well as the different elastic energy terms, are compared for one and two torons and for a twistion when normalized by the corresponding energies of the unwound homeotropic state. For $d/p=0.85$ and elastic constants of 5CB (Table II, both the toron and the twistion are metastable states while the unwound cholesteric LC is the ground-state with free energy per unit volume $2\pi^2k_{22}/p^2$ with respect to which the other energies were normalized. The free energies of the twistion and toron structures were calculated using Eq. \ref{eq:fe_handedness} for the entire computational volume described above and for the elastic constants of 5CB (Table I). The total free energy was then normalized to the energy of an equivalent volume of unwound uniform director. The free energy of a twistion is larger than that of the unwound state and the toron configuration, suggesting that the twistions are metastable states for the used material parameters and cell confinement, albeit such structures could potentially become equilibrium states under different conditions. Overall, we envisage the possibility of existence of a larger family of twistions in which even numbers of self-compensating point defects (or small disclination loops) can be stabilized by different configurations of the stretched loops of double twist tube interspacing them. Such structures will be explored both experimentally and numerically in our future studies. 

\vspace{10mm}
Table II. Free energy of twistions and torons normalized to the corresponding total free energy of the unwound uniform state of the confined cholesteric LC $(F_0)$. Material parameters for 5CB (Table I) were used to obtain these data.
\begin{center} 
\begin{tabular}{lccc}
\hline \hline
         & $One~toron$ & $Two~torons$ & $Twistion$\\  \hline
	$F$         & 1.0616 & 1.1232 & 1.1477 \\
	$F_{11}$ & 0.0366 & 0.0731 & 0.0713 \\
	$F_{22}$ & 0.9044 & 0.8098 & 0.8688 \\
	$F_{33}$ & 0.1284 & 0.2557 & 0.2273 \\
	$F_{24}$ & -0.0077&-0.0154&-0.0198 \\
\hline \hline
\end{tabular}
\end{center}

By examining the structure of point defects, we find that all of them are of the hyperbolic type, both in the case of twistions and torons. However, only some of the point defects are accompanied by handedness reversal. For example, out of the four point singularities of the twistion, only three exhibit this reversal behavior (Fig. \ref{handednessFig4}). Unlike the other three that are spaced between the highly twisted interior of the twistion and the untwisted surrounding and have symmetry axes along the cell normal, the fourth hyperbolic defect has an in-plane symmetry axis and is embedded mainly within the twisted region (Fig. \ref{handednessFig4}). This explains the observation that the twist handedness in its vicinity remains consistent with that of the ground state chiral nematic LC, unlike in the cases of the other 3 hyperbolic defects of the twistion. Point defects of torons always exhibit the handedness reversal in their vicinity (Fig. \ref{handednessFig5}), also consistent with the fact that this handedness reversal helps to match the main part of the twisted localized structure with the unwound LC background.

\section{Conclusions}

To conclude, contrary to common expectations, we have demonstrated that locally measured handedness of twist in chiral LCs can reverse and be opposite to that of the equilibrium state. Our analysis shows that this energetically costly twist reversal occurs to match the solitonic structure with the optimal localized twisted configuration to the unwound surrounding far-field director. Since chiral condensed matter systems, which include chiral LCs, are considered to be promising candidates for the practical realization of various 2D and 3D twisted solitons, such as skyrmions and Hopfions, our findings may have broad implications for the stability of such structures and their experimental realization. In addition to the confinement of flat homeotropic cells, similar effects of handedness reversal can occur in other geometries, such as that of spherical or complex-topology high-genus cholesteric LC droplets, cylindrical capillaries, etc. \cite{Tortora2011PNAS, Sec2014NatComm, jeong2015pnas, Nayani2015, Beller2014}. In many LCs with low-value twist elastic constants, such as chromonic LCs \cite{Tortora2011PNAS, jeong2015pnas, Nayani2015}, director twist often occurs to replace more energetically costly splay and bend deformations. While handedness directionality is not pre-determined by the intrinsic molecular chirality in these LCs, we foresee a rich variety of problems that can be addressed by applying the pseudotensorial handedness analysis \cite{Efrati2014} in the probing of LCs under different forms of confinement, which we have applied here to only a small subset of problems benefiting from it. Quantitative characterization of 3D patterns of twist handedness will complement and expand the existing visualization approaches that have shown to be valuable tools in the characterization of LC structures and defects \cite{Copar2013lc}.

\section{Acknowledgments}
~~~ This work was supported by NSF through Grant No. DMR-1410735. We thank Efi Efrati, A. Hess, H. Osman, and M. Tasinkevych for discussions. P.J.A. and I.I.S. thank the Boulder Condensed Matter Summer School, during which this work was initiated and emerged from stimulating discussions.

\end{multicols}

\end{document}